# ELECTROOCULOGRAPHY DATASET FOR OBJECTIVE SPATIAL NAVIGATION ASSESSMENT IN HEALTHY PARTICIPANTS

Mobina Zibandehpoor[1]    Fatemeh Alizadehziri[1]    Arash Abbasi Larki[1]    Sobhan Teymouri[1]
Mehdi Delrobaei[1,*]

[1]Mechatronics Department, Faculty of Electrical Engineering, K. N. Toosi University of Technology, Tehran, Iran
[*]Corresponding author: delrobaei@kntu.ac.ir

## ABSTRACT

In the quest for understanding human executive function, eye movements represent a unique insight into how we process and comprehend our environment. Eye movements reveal patterns in how we focus, navigate, and make decisions across various contexts. The proposed dataset includes electrooculography (EOG) signals from 27 healthy subjects, capturing both vertical and horizontal eye movements. The recorded signals were obtained during the video-watching stage of the Leiden Navigation Test, designed to assess spatial navigation abilities. In addition to other data, the dataset includes scores from the Mini- Mental State Examination and the Wayfinding Questionnaire. The dataset comprises carefully curated components, including relevant information, the Mini-Mental State Examination scores, and the Wayfinding Questionnaire scores, encompassing navigation, orientation, distance estimation, spatial anxiety, as well as raw and processed EOG signals. These assessments contribute more information about the participants' cognitive function and navigational abilities. This dataset can be valuable for researchers investigating spatial navigation abilities through EOG signal analysis.

## 1    Background & Summary

Executive functions encompass cognitive processes and mental abilities, including working memory, cognitive flexibility, and inhibition control. These functions are instrumental in planning, monitoring, and effectively achieving our goals. They are pivotal in managing daily tasks, problem-solving, and adapting to novel situations. Executive functions intersect with decision-making, long-term memory, and environmental adaptation, particularly in the context of spatial navigation Horowitz-Kraus et al. [2023], Marzocchi et al. [2020].

Therefore, spatial navigation is a multifaceted cognitive function essential for planning and finding routes in one's environment. It mainly encompasses the awareness of one's current position, orienting oneself in space, identifying the goal location, and formulating a navigational path linking these points. Psychometric test scores have been shown to predict an individual's performance on dynamic spatial navigation tasks by assessing cognitive abilities related to these processes Garg et al. [2024], Laczó et al. [2021], Korthauer et al. [2017], Wei et al. [2020].

Navigation skills significantly impact our daily lives and professional endeavors since they comprise a range of skills such as wayfinding, mental mapping, and spatial orientation der Kuil et al. [2022], Prevratil et al. [2023]. Recent research suggests that professional athletes, including soccer players rely on their mind-guiding system and cognitive map to enhance physical performance. This finding highlights the importance of navigation ability in optimizing motor skills and preventing falls, even among non-sporting individuals or patients undergoing vestibular system rehabilitation Gamba et al. [2022].

From cognitive map development to taking advantage of navigation aids, people use spatial navigation to discover their way through unknown locations and new places using maps, GPS, or verbal instructions. Besides, spatial navigation abilities are important for finding joy in discovery and managing spatial anxiety Muffato et al. [2022]. Assessing spatial navigation is crucial since it allows monitoring of various conditions, such as Alzheimer's disease, where changes in direction perception occur even in the early stages. Additionally, evaluating this skill is vital for measuring age-related declines in navigational abilities Smith et al. [2024]. Many tools and techniques have been designed to assess spatial navigation skills. These tools include eye tracking Keskin et al. [2019], electrooculography (EOG) Postelnicu et al. [2012], and electroencephalography (EEG) Miyakoshi et al. [2021]. Additionally, virtual navigation tests Lin et al. [2009], Roth et al. [2020] and digital games like Sea Hero Quest offer insights into visuospatial processing, memory, and executive functions, among other aspects of cognition Garg et al. [2024]. Knowledge about spatial navigation and its assessment helps identify early signs of cognitive impairments, assess hippocampal functioning, and develop approaches to improve deficient navigational behavior, especially in the elderly and those with neurological diseases Rekers et al. [2024]. The EOG measures eye movements through extra-ocular muscle electrical signals. It is nonintrusive and ideal for continuous field use. It can precisely measure eye movements and find applications in brain-computer interfaces Moon et al. [2023], Nezvadovitz et al. [2022]. While EOG has a lower spatial resolution than eye tracking, it excels in capturing temporal measures. Notably, EOG can assess parameters like Quiet Eye duration and spectral decomposition, which aid in distinguishing between low-frequency oscillations and evoked responses over time Gallicchio et al. [2024]. Furthermore, EOG signals can be processed in real-time using model-oriented denoising approaches, preserving the inherent characteristics of the signals without distortion, unlike traditional methods such as bandpass filtering Gunawardane et al. [2024], Ahmed et al. [2022].

Utilizing engineering tools for data collection in cognitive psychology can significantly enhance relevant research by yielding more detailed and profound insights. This approach leads to increased efficiency, improved data quality, and, in some instances, reduced participant burden. Existing literature highlights that engineering tools offer cost-effective ways to enhance data quality Rhemtulla et al. [2012]. Moreover, such tools facilitate the conversion of data types between devices and applications, thus ensuring seamless interpretation and analysis Reymond et al. [2013]. Moreover, they increase the reliability and validity of collected information and reduce participant fatigue and practice effects, thereby enhancing overall research outcomes and insights into cognitive processes Rhemtulla et al. [2012].

Recent progress in EOG technology has led to various approaches for capturing and analyzing eye movements. A new EOG system using an ATmega AVR microcontroller to capture vertical and horizontal eye movements has been developed. This system integrates dual-channel filtering with High-Pass and Low-Pass Filters to manage the signal range of 0.1 to 10 Hz. Tested with diverse volunteers, this setup effectively gathers eye movement data, demonstrating potential applications in rehabilitation and assistive technologies Abdel-Samei et al. [2023]. Additionally, another study offered a dataset of EEG and EOG recordings from four patients with advanced locked-in syndrome (LIS) due to amyotrophic lateral sclerosis (ALS). The data, collected across several visits, contributes valuable insights into the progression of ALS and supports the development of assistive technologies and brain-computer interfaces (BCIs), enhancing clinical management and therapeutic strategies Jaramillo-Gonzalezl et al. [2021]. Researchers also investigated EOG while participants tracked a moving cross that shifted every 1,250 ms. The cross's horizontal displacements ranged from 1 to 7 degrees, with occasional vertical shifts. Participants were also instructed to blink when the cross changed to a circle. This setup provided calibrated data on the EOG signal's response to different gaze shifts and blinks Reichert et al. [2020]. Another research employed the g.tec USBamp biosignal amplifier, sampling at 256 Hz, to record EOG data. Their methodology included bandpass filtering from 0 to 30 Hz and a 50-Hz notch filter. The study involved recording eye movements from six healthy individuals, focusing on 600 saccades and 300 blinks per participant using a conventional electrode arrangement, including horizontal and vertical electrodes and ground and reference electrodes positioned on the forehead and behind the left ear Barbara et al. [2020]. However, existing EOG datasets have largely overlooked cognitive assessment. The primary focus has been on technical performance and practical applications rather than evaluating cognitive abilities. In contrast, this study aims to fill this gap by utilizing the EOG device to assess cognitive functions, exploring how eye movement patterns can provide insights into human spatial navigation ability.

Eye movement datasets for effectively evaluating spatial navigation are currently limited and not widely available. Existing datasets primarily rely on eye tracker devices. However, there is a need for more diverse datasets to assess spatial navigation. EOG devices offer several advantages over traditional eye trackers. Unlike eye trackers, which require a direct line of sight to the eye, EOG measures electrical signals generated by eye movements through electrodes placed around the eyes. This allows EOG devices to function without requiring direct visual contact. Additionally, EOG devices perform well under various lighting conditions, including bright light, darkness, and rapid changes in lighting. In contrast, eye trackers may lose accuracy in poor lighting conditions. EOG devices offer a straightforward design, making them cost-effective and minimizing power consumption. These attributes make EOG devices particularly well-suited for applications with resource constraints. Moreover, their compact size and portability enable their use in various environments. The proposed dataset comprises EOG signals collected from healthy and young subjects. These signals were recorded during the video-watching phase of a previously validated navigation assessment, the Leiden Navigation Test (LNT) van der Ham et al. [2020], designed to evaluate spatial navigation abilities. Additionally, the dataset includes scores from questionnaires, such as the Mini-Mental State Examination (MMSE) Tombaugh et al. [1992] and the Wayfinding Questionnaire (WQ) De Rooij et al. [ 2019]. We hope researchers investigating spatial navigation abilities through EOG signal analysis find this dataset valuable.

## 2 Methods

This section overviews the study methodology, including participant details, equipment specifications, task execution protocols, and data preprocessing procedures applied to the collected EOG signals.

### 2.1 Participants

Twenty-seven healthy university students, comprising 14 males and 13 females, with a mean age of 21.78 ± 1.59, participated in this study. This data was collected in the Mechatronics Laboratory, K. N. Toosi University of Technology. All the participants reported no known diseases or visual impairments. Those who required glasses were instructed to wear them during the tasks. All participants received the required information about the study, and to safeguard their privacy, a unique ID number was assigned to each participant. Informed consent was obtained from all participants, and the study strictly adhered to the principles outlined in the Declaration of Helsinki.

### 2.2 Equipment

A two-channel EOG headband (Zehnafzar Rayan Co., Isfahan, Iran) was employed to record the eye movement signals with high precision. The first channel captures vertical eye signals in this device, while the second records horizontal eye signals. The device functions at a sampling rate of 250 Hz, producing signals within the voltage range of ±2.4 volts. It includes five gold cup electrodes integrated into flexible material. According to Fig. 1, two electrodes are positioned near the temples to monitor horizontal eye movements, while two more electrodes are placed above and below one eye to capture vertical movements. A fifth electrode, serving as the reference, is located on the opposite side of the forehead.

The device operates wirelessly, is powered by a rechargeable battery, and connects to a PC via Bluetooth. This wireless link enables real-time signal display and automatic recording through a dedicated MATLAB® software, ensuring efficient and accurate data c ollection. The headband's design and connectivity allow uninterrupted eye movement monitoring, enhancing the collected dataset's robustness.

### 2.3 Protocols

All participants completed the informed consent form, ensuring they understood the study's procedures and agreed to participate. They then filled out the WQ, which assessed their navigational strategies and preferences. The examiner then administered the MMSE to evaluate cognitive function. All the questionnaires and forms were translated into Persian.

Participants were seated on a chair positioned 2.15 meters away from a 46-inch monitor, as shown in Fig. 2. The EOG headband, with electrodes coated in TEN20 conductive gel, was carefully fitted onto the participant's head. Special attention was given to ensure the electrodes made proper contact with the skin, avoiding interference from hair. The headband was powered on and automatically connected to the examiner's computer via Bluetooth, ensuring seamless data transmission.

Once the setup was complete, participants were briefed on the procedure, and the LNT video segment was initiated. This phase involved watching a series of animated navigational tasks designed to assess spatial awareness and navigation skills. Upon completing the video segment, participants were given a wireless mouse to answer the image questionnaire related to the LNT. This offered insights into their perception and understanding of the navigational tasks.

Finally, the test results were recorded under the participant's assigned ID to ensure confidentiality and accurate data tracking. The EOG headband was then carefully removed.

### 2.4 Data Processing

The proposed methodology, referred to as Algorithm 1, utilizes advanced signal processing techniques to detect ocular events, specifically b links, saccades, and fi xations, across both horizontal and vertical electrooculography (EOG) channels. The algorithm is structured into six primary stages, each aimed at enhancing the precision and robustness of eye movement detection.

**Algorithm 1** EOG Signal Processing
───────────────────────────────────────────
    **Input:** verticalEOG, horizontalEOG    **Output:** compositeLabel
    **A. Preprocessing:**
vEOG, hEOG ← ApplyMovingMedianFilter(verticalEOG), ApplyMovingMedianFilter(horizontalEOG)
vEOG, hEOG ← ApplyBandpassFilter(vEOG), ApplyBandpassFilter(hEOG)
    **B. Blink Detection:** vBlinkLabel ← DetectBlinks(vEOG)
    **C. Saccade Detection:**
vSaccadeLabel, hSaccadeLabel ← DetectSaccades(vEOG, vBlinkLabel), DetectSaccades(hEOG)
    **D. Saccade Duration Thresholding:**
vSaccadeLabel, hSaccadeLabel ← ApplySaccadeDurationThreshold(vSaccadeLabel, hSaccadeLabel)
    **E. Label Fusion:** compositeLabel ← FuseLabels(vBlinkLabel, vSaccadeLabel, hSaccadeLabel)
    **F. Fixation Identification and Thresholding:**
compositeLabel ← IdentifyAndThresholdFixations(compositeLabel)
**return** compositeLabel
**Function DetectBlinks(vEOG):**
peaks ← FindPeaksAboveThreshold(vEOG, blinkThreshold)
blinks ← [ ]
**for** each peak in peaks **do**
    onset, offset ← FindOnset(vEOG, peak), FindOffset(vEOG, peak)
    blinks.append((onset, offset))
**end for**
**return** CreateLabelSignal(blinks, 3)
**Function DetectSaccades(EOG, blinkLabel):**
derivative ← CalculateDerivative(EOG)
saccades ← [ ]
**for** each region in derivative above threshold **do**
    **if** blinkLabel is not None **and** region overlaps with blink **then**
        **continue**
    **end if**
    onset ← FindZeroCrossing(derivative, region.start, 'backward')
    offset ← FindZeroCrossing(derivative, region.end, 'forward')
    saccades.append((onset, offset))
**end for**
**return** CreateLabelSignal(saccades, 2)
**Function ApplySaccadeDurationThreshold(saccadeLabel):**
**for** each saccade in saccadeLabel **do**
    **if** Duration(saccade) < minimumSaccadeDuration **then**
        RemoveSaccade(saccadeLabel, saccade)
    **end if**
**end for**
**return** saccadeLabel
**Function FuseLabels(vBlinkLabel, vSaccadeLabel, hSaccadeLabel):**
**return** MaximumFunction(vBlinkLabel, vSaccadeLabel, hSaccadeLabel)
**Function IdentifyAndThresholdFixations(compositeLabel):**
potentialFixations ← FindRegionsBetweenEvents(compositeLabel)
**for** each fixation in potentialFixations **do**
    **if** Duration(fixation) ≥ minimumFixationDuration **then**
        LabelRegion(compositeLabel, fixation, 1)
    **else**
        LabelRegion(compositeLabel, fixation, 0)
    **end if**
**end for**
**return** compositeLabel
───────────────────────────────────────────

The initial stage encompasses preprocessing vertical and horizontal EOG signals by applying a moving median filter for smoothing, followed by a bandpass filter. These preprocessing steps are designed to attenuate noise and eliminate low-frequency drift, effectively improving the signal-to-noise ratio and facilitating subsequent analytical processes. Algorithm 1 focuses predominantly on the vertical EOG channel for blink detection, identifying potential blinks by pinpointing peaks that surpass a predefined threshold. The derivative of the signal is scrutinized to accurately ascertain the onset and offset of each blink, thereby enabling precise temporal localization of blink events. Detected blinks are marked with a value of 3 in the vertical label signal.

The detection of saccades is conducted independently on both vertical and horizontal channels. The algorithm employs thresholds on the derivative of the EOG signal, with the onset and offset of saccades identified through zero-crossing analysis of this derivative. Importantly, the algorithm for detecting saccades in the vertical channel excludes regions previously identified as blinks, thereby minimizing the incidence of false positives. Detected saccades are designated a value of 2 in their respective label signals.

Following the initial detection phase, a minimum duration threshold is applied to the identified saccades in both channels to differentiate genuine saccades from potential artifacts or microsaccades. Events that meet or exceed this duration threshold retain their value of 2 in the label signal, while those falling short are disregarded in subsequent analyses. The vertical and horizontal label signals are integrated using a maximum function, yielding a composite label signal that comprehensively represents eye movement events across both channels. In overlapping events, the algorithm prioritizes blink detection (value 3) over saccade detection (value 2) to ensure clear delineation of eye movementsHolmqvist et al. [2011].

The final stage involves identifying and thresholding fixations initially recognized as periods between detected saccades and blinks. A minimum duration threshold is applied to these potential fixations, with those that meet or exceed the threshold assigned a value of 1 in the composite label signal, while those that do not are categorized as unknown events and designated a value of 0. The resulting composite label signal provides a comprehensive representation of eye movement events, classified as follows: 3 for blinks, 2 for saccades, 1 for fixations, and 0 for unknown or ambiguous events. This methodological approach offers a robust framework for EOG signal analysis, yielding detailed insights into eye movement patterns. The systematic design ensures efficient processing and clear delineation of eye movements, even in contexts characterized by overlapping events. Furthermore, integrating duration thresholds and post-processing strategies enhances accuracy while mitigating false positives, providing a comprehensive analysis.

## 3 Data Records

The data comprises several components, succinctly illustrated in Fig. 3. The files containing the raw data and the codes for data analysis are available at figshare.comZibandehpoor et al. [2024].

- Dataset.csv: This section encompasses an Excel file that details pertinent information about each of the 27 participants. The dataset includes age, gender, educational level, and MMSE scores. Additionally, it provides detailed scores for each of the subcomponents of the WQ. Similarly, it includes scores for the LNT image questionnaire. Table I presents the range and definitions for each abbreviated concept, which is crucial for interpreting the results accurately.

  The MMSE is a cognitive assessment tool used to screen for impairments and monitor changes over time. It measures cognitive domains like orientation, memory, attention, language, and visuospatial abilities. The MMSE includes recalling objects, naming items, and copying shapes. Scoring ranges from 0 to 30, with scores from 24 to 30 indicating normal cognition, 18 to 23 suggesting mild impairment, 10 to 17 indicating moderate impairment, and below 10 denoting severe impairment. In the given data collection, all participants scored above 27, indicating minimal to no cognitive impairment.

  The WQ is designed to evaluate three distinct subscores of navigational skills: navigation and orientation, distance estimation, and spatial anxiety. Each category is measured using a series of questions based on the Likert scale, which ranges from 1 to 7. In this scale, a score of 1 represents the lowest possible state, indicating minimal proficiency or confidence, while a score of 7 denotes the highest level of proficiency or confidence. Specifically, the navigation and orientation subscore is assessed with 11 questions, distance estimation with three questions, and spatial anxiety with eight questions. The range for each subscore is clearly defined based on the number of questions and the Likert scale utilized, ensuring a detailed and refined understanding of the participant's capabilities and perceptions in these areas.

Additionally, LNT focuses on the participant's ability to recognize and interpret landmarks and paths. These subscores are divided into five categories: landmark recognition, path: survey, location: egocentric, location: allocentric, and path: route. The number of questions for each category is carefully balanced, with 8 questions for landmark recognition and 4 questions each for the remaining categories. The scoring system for the LNT component is straightforward: participants receive a score of 1 for each correct response and 0 for each incorrect response. Because all the forms and questionnaires were translated into Persian, we discovered that the location egocentric subscore had been translated in a way that changed its original meaning. As a result, we decided to remove this subscore from the dataset.

The file includes several features designed for easy access and use. These features cover vital metrics; it provides a summary of extracted features from each EOG signal, detailing crucial metrics like the number of blinks, the number of saccades (rapid eye movements), summation and average saccade durations, the number of fixations, and the summation and average fixation durations for each participant.

- LNT Image Questionnaire Answer.csv: This section features an organized Excel file that captures participants' responses to the LNT image questionnaire questions. In addition to the participants' answers, the file includes the correct answers for each question, facilitating a comparison and analysis of the participants' performance. This structured data allows for an in-depth examination of individual and group accuracy in identifying and interpreting the LNT landmarks, thereby contributing valuable insights into spatial cognition and navigational abilities.

- EOG Data: This section is organized into two parts. The first part contains a folder filled with raw EOG signals recorded from each participant as they watched the LNT video segment. These files are saved in the .mat format, ensuring compatibility with various versions of MATLAB® software. Each file is labelled with the participant's unique identification number. Upon opening these files, researchers will find horizontal signals labelled as A and vertical signals labelled as B, reflecting the eye movements recorded during the video. The signal span corresponds precisely to the video's length, providing a complete and uninterrupted record of the eye movement data for that period.

The second part of this section comprises Excel files that offer a comprehensive summary of features extracted from each participant's EOG signals. These files contain key metrics, including the total count, amplitude, and duration of blinks; the total count, amplitude, and duration of saccades (rapid eye movements); and the total count, amplitude, and duration of fixations (periods of eye stability). These measurements are reported for vertical, horizontal, and concatenated EOG signals, providing a detailed analysis of oculomotor behavior across all signal dimensions. They are invaluable for analyzing the participants' visual attention and eye movement patterns in response to the LNT video segment. The nature of this data allows for in-depth studies into visual and cognitive processing during navigation tasks, offering rich insights into how individuals perceive and interact with the environment.

# 4 Technical Validation

In this study, we curated an EOG dataset using a dual-channel headband. The EOG signals were recorded at a sample rate of 250 Hz, ensuring a fine temporal resolution for capturing eye movement dynamics. The system employed a wide dynamic range, enabling precise detection of subtle and pronounced eye movements. The utilized analog-to-digital converter had a high resolution of 24 bits, facilitating accurate digitization of the analog EOG signals.

Our recording device incorporated low-pass and high-pass filters during data collection to enhance data quality and maintain data integrity. Specifically, the low-pass filter was configured with a cut-off frequency of 20 Hz to eliminate high-frequency noise that might otherwise contaminate the EOG signals. Simultaneously, the high-pass filter's cut-off frequency was set at 0.05 Hz to remove slow drifts and baseline shifts. This filtering process preserved the essential characteristics of eye movements while effectively minimizing the impact of irrelevant noise and artifacts.

To enhance the reliability of the EOG recordings, we conducted calibration sessions before each recording. During these sessions, the system was adjusted to minimize measurement error. Following calibration, validation sessions were performed to verify the system's accuracy, with recording commencing only once the measurement error was within acceptable limits. This calibration and validation process ensured the data were accurate and reliable. The recording system's noise characteristics were also carefully managed. The input short circuit noise was measured at 4 µV, indicating a high-quality signal with minimal interference. Additionally, the system's least significant bit (LSB) corresponded to 2.86 µV, allowing for detecting small changes in the EOG signal. This high resolution enhances the accuracy of the recorded data.

We implemented several measures to maintain high data quality throughout the data recording process. Notably, the impedance of the electrodes was continuously monitored to ensure it remained within acceptable levels. This proactive approach significantly reduced artifacts and improved signal quality. The recorded data were also inspected for artifacts, such as those caused by muscle movements or blinks, and these artifacts were either corrected or marked for exclusion in subsequent analyses. These steps, particularly the continuous monitoring of electrode impedance, ensured the highest possible quality and reliability of the EOG data.

The dual-channel EOG data were recorded with high temporal and voltage resolution, low noise levels, and rigorous calibration and validation procedures. Including both low-pass and high-pass filtering during data collection further ensured the high quality and reliability of the EOG data presented in this study.

## 5 Usage Notes

Data collection was performed utilizing Matlab R2023b. The data is also executable and verified for compatibility and functionality across other MATLAB versions.

## 6 Code Availability

The files containing the codes for data analysis are available at github.comAra [2024].

## 7 Acknowledgements

We gratefully acknowledge the participation of all study participants.

## Author Contributions Statement

M.Z. conceived the experiments, and M.Z., F.A., and A.A. wrote the manuscript. A.A. and S.T. handled data processing, debugging, and graph preparation. M.D. managed the project and provided essential revisions to the paper. All authors reviewed and approved the final manuscript and are responsible for all elements of the work.

## 8 Competing Interests

The authors declare no competing interests.

## 9 Figures & Tables

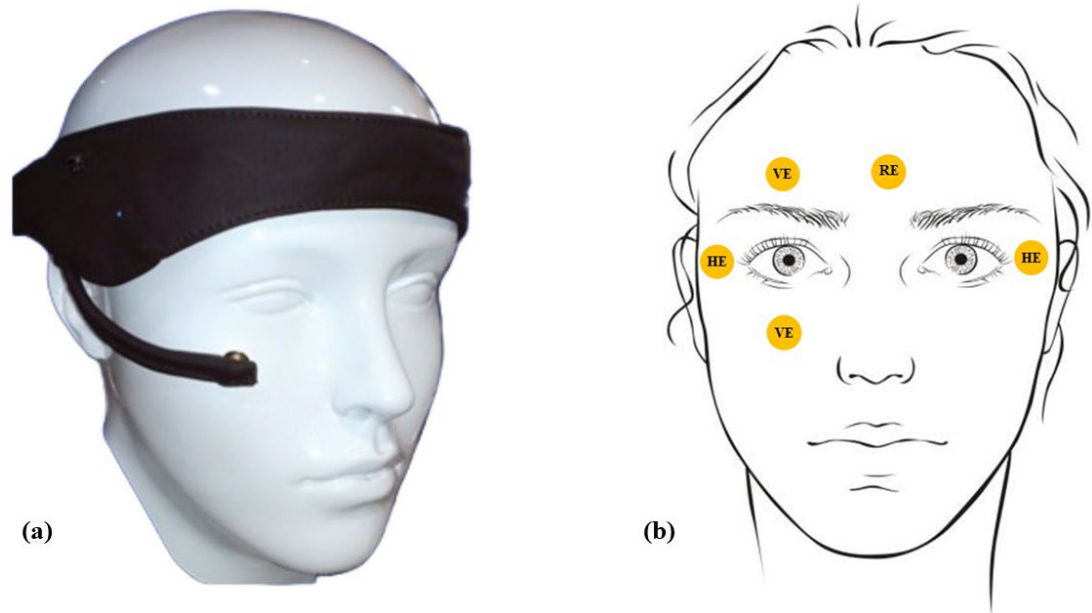

Figure 1: (a) The placement of the headband on the participant's head. (b) The positioning of each electrode on the EOG headband: VE = Vertical Electrode, HE = Horizontal Electrode, RE = Reference Electrode.

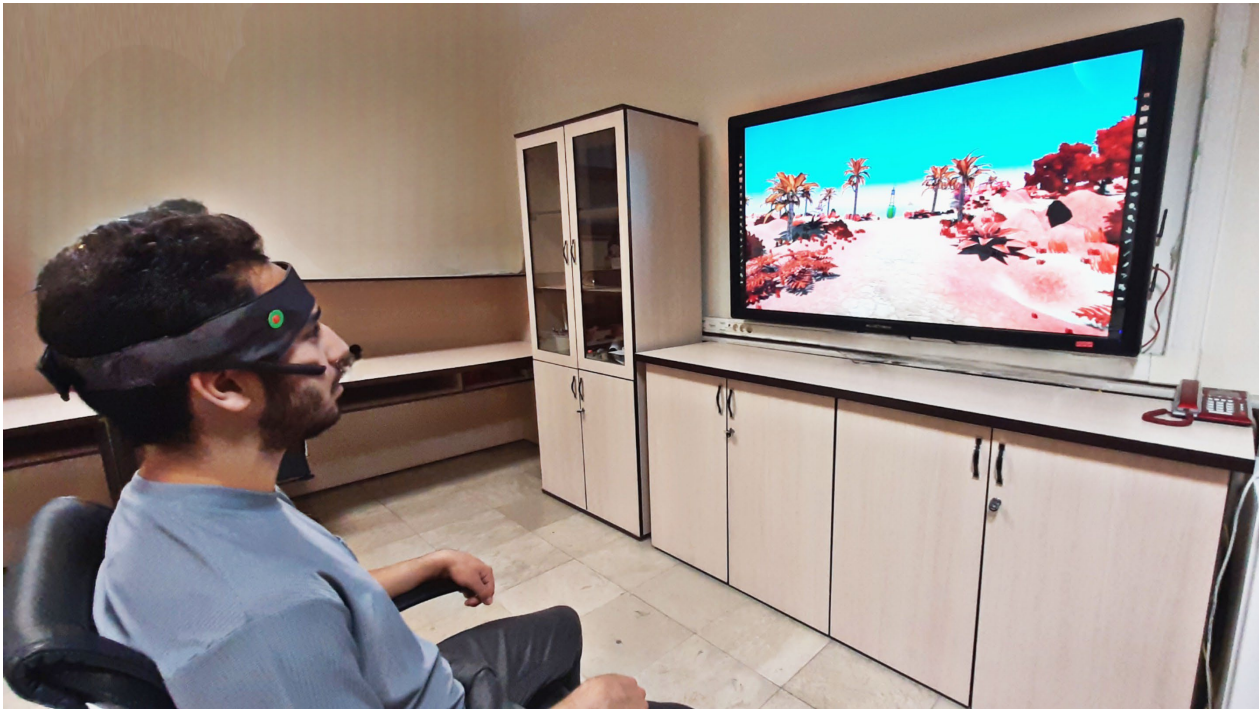

Figure 2: The participant wearing the EOG headband is sitting in front of the monitor.

Table 1: Summary of demographic and range of variables.

| Row Names | Range and Concepts |
|---|---|
| Gender | M = Male, F = Female |
| Age | $19 \leq$ Age $\leq 25$ |
| Education | B = Bachelor, M = Master |
| MMSE Score | $0 \leq$ MMSE $\leq 30$ |
| Navigation and orientation (WQ) | $11 \leq$ Navigation and orientation $\leq 77$ |
| Distance estimation (WQ) | $3 \leq$ Distance estimation $\leq 21$ |
| Spatial anxiety (WQ) | $8 \leq$ Spatial anxiety $\leq 56$ |
| Landmark recognition (LNT) | $0 \leq$ Landmark recognition $\leq 8$ |
| Path survey (LNT) | $0 \leq$ Path survey $\leq 4$ |
| Location allocentric (LNT) | $0 \leq$ Location allocentric $\leq 4$ |
| Location egocentric (LNT) | $0 \leq$ Location egoocentric $\leq 4$ |
| Path route (LNT) | $0 \leq$ Path route $\leq 4$ |

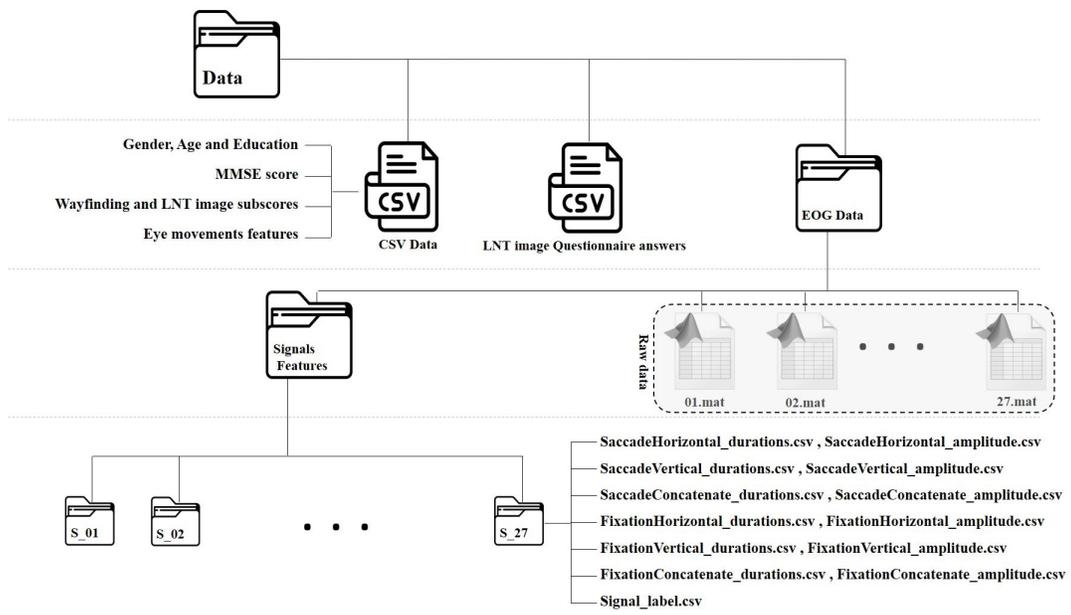

Figure 3: Raw data folder structure.